\documentclass[twocolumn]{aastex62}
\pdfoutput=1 
\usepackage{amsmath,amstext}
\usepackage[T1]{fontenc}
\usepackage{apjfonts} 
\usepackage[figure,figure*]{hypcap}
\usepackage{multirow}
\usepackage{graphicx}
\graphicspath{{figs/}}
\newcommand{\comment}[1]{}

\begin{document}
\title{Variability of magnetically-dominated jets from accreting black holes}
\author{Agnieszka Janiuk}
\author{Bestin James}
\author{Ishika Palit}
\affiliation{Centrum Fizyki Teoretycznej PAN
Al. Lotnik\'ow 32/46, 02-668 Warsaw, Poland}

\begin{abstract}
Structured jets are recently invoked to explain the complex emission of gamma ray bursts, such as GW 170817. Based on the accretion simulations, the jets are expected to have a structure that is more complex than a simple top-hat. 
Also, the structure of launching regions of blazar jets should influence their large scale evolution. This is recently revealed by the interactions of jet components in TXS 0506+056, where the jet is observed at a viewing angle close to zero.

Observational studies have also shown an anti-correlation between the jet variability, measured e.g. by its minimum variability time scale, and the Lorentz factor, that spans several orders of magnitude and covers both blazars and GRBs samples. 
Motivated by those observational properties of black hole sources, we investigate the accretion inflow and outflow properties, 
by means of numerical GR MHD simulations. 
We perform axisymmetric calculations of the structure and evolution of central engine, composed of magnetized torus around Kerr black hole that is launching a non-uniform jet. We probe the jet energetics at different points along the line of sight, and we measure the jet time variability as localized in these specific regions. We quantify our results by computing the minimum variability timescales and power density spectra. We reproduce the MTS-$\Gamma$ correlation and we attribute it to the black hole spin as the main driving parameter of the engine. We also find that the PDS slope is not strongly affected by the black hole spin, while it differs for various viewing angles.

\keywords{accretion, accretion disks, black hole physics, magnetic fields, hydrodynamics, MHD, gamma rays: bursts, galaxies: jets}
\end{abstract}

\section{Introduction}
The fastly variable accretion flows are found in a number of different types of astrophysical black hole sources. At largest scales, they are present in the cores of active galaxies. 
In the radio-loud objects, such as blazars, the variability of the inflow can be transmitted to the outflow properties. In these sources, 
the relativistic jets are pointing to our line of sight. 
In addition, many similarities are found between the jet physics in blazars and in gamma ray bursts. 
The latter are observed from extragalactic distances, but 
operate at smaller scales, within the stellar-mass accreting black holes and in collapsing star's environment.
Blazars and gamma ray bursts share the properties of their jets, despite different Lorentz factors and accreting black hole masses \citep{Wu2016}.
Launching and collimation mechanisms are common: thick disk or corona, pressure gradient in surrounding wall, external (matter dominated) jet, or 
toroidal magnetic field.
Acceleration of jets occurs due to both magnetic field action
field and accretion disk rotation (see \cite{Fragile2008} for a review).
The blazar jets are
 Poynting-dominated, and powered by the Blandford-Znajek mechanism which can extract energy from a rotating black hole. This mechanism is now 
well known and tested in the purpose of a jet launching, but observations are showing variability in the jet emission. Multiple shocks that collide in the jet, 
can lead to multiple emission episodes and can account for the fluctuating light curve of the gamma ray bursts \citep{1997Kobayashi}. A reasonable interpretation of this effect is that the variability 
observed in the jets can directly reflect the central engine variability. The latter is tightly related to the action of magnetic fields in the center of the 
galaxy, or in the GRB central engine.
What is more, the structure of a jet at its base is possibly much more complex than a simple top-hat and can be revealed by the afterglow observations and interactions of the large scale jet with the surrounding medium, e.g. with the post-merger wind in GW 170817 \citep{Urrutia2021}. 
Even though the observed lightcurves and spectra are primarily the result of the jets interaction with the circumburst medium, the initial structure of the jet at its base is also affecting the final emission.
Also, interactions are possible between the different components of precessing blazar jets, such as in TXS 0506+056 \citep{2019Britzen} .

The variable energy output from the central engine implies the varying jet Lorentz
factor, as shown e.g. by \cite{Sap2019}. This may lead to occurrence of internal shocks, and affect both GRBs and blazars
observed variability \citep{2008Begelman, 2016Bromberg}. 
Unification of the models accross the black hole mass scale, from
GRBs to blazars, is not straightforward though. The most uncertain aspect is whether
the magnetically arrested disk (MAD) state drives the jets in both type of sources, or rather halts the GRB emission, as studied by \cite{Lloyd-Ronning2018}.
In the MAD mode, the 
flux accumulated at the BH horizon, and the interchange
instability rather than magneto-rotational instability MRI governs the minimum timescale of variability
\citep{2011Tchekhovskoy, 2014Tidal}.
In contrast, in the SANE mode, the MRI dives variability of the jets, as directly related to the accretion variability timescales \citep{2013Penna, 2019Porth}. Finally, the blazar disks are subject to a different physical conditions than the GRB disks, and in the latter, thermal instabilities of the neutrino-dominated accretion flows may play a role, triggering also the episodic jet jections \citep{2010Janiuk, 2014Cao}

Here we explore the scenario of magnetically driven accretion and jet variability related to the MRI timescale. We confirm the existence of
the correlations between the 
inferred minimum variability timescale and magnetic field strength as well as the black hole spin, as expected for the Blandford-Znajek driven jets.
We compare the resulting timescales with observed ones, taken from the sample by \cite{Wu2016} and conclude that they represent well both classes of sources, namely blazars and GRBs.

\section{Model}

We present here the two-dimensional magneto-hydrodynamical models computed in full General Relativity (general relativistic magneto-hydrodynamics, GR MHD). The numerical scheme is our implementation 
of the code HARM (\cite{Gammie2003, Janiuk2013, Sap2019} ).
Our initial condition assumes the existence of a pressure equilibrium torus, embedded in the poloidal magnetic field 
(Fig.\,\ref{fig1}, top panel). 
 Such 2D studies of a compact magnetized tori around black holes have been performed already by different groups \citep{mckinney12, fernandez2013,sadowski2015,quian17}.  
Our approach is based on similar methodology, while the focus of the present study is given to measuring the variability of the jet. The novel aspect of our analysis is that we consider a structured jet morphology and we attempt to compare our results with some observables. 

The jet launched from the central engine is powered by rotating black hole and mediated by magnetic fields.
The Kerr black hole accretes matter from the torus, and its rotation affects the magnetic field evolution. 
The models are parameterized with the black hole spin, and the initial magnetization of the matter.
Code works in GR framework, so dimensionless units are adopted, with $G=c=M=1$. Hence, geometrical time is given as
$t=GM/c^{3}$, where M is the black hole mass.
In this way, we are able to model the launching and variability of jets in both supermassive black hole environment, and in gamma ray bursts.

The chosen configuration of the torus structure is that of \cite{Chakrabarti1985}. Here, the angular momentum distribution has a power law relation with the von Zeipel parameter $\lambda = (l/\Omega)^{1/2}$, where $l$ denotes the specific angular momentum and $\Omega$ denotes the angular velocity. The size of the torus is fixed in geometrical units, and its inner radius is located at $r_{\rm in} = 6 r_{g}$, its density maximum is located  at $r_{\rm max}= 16.5 r_{g}$, and the outer edge is at at about $r_{\rm out}= 40$.

We embed the initial torus in poloidal magnetic field, that was proven to drive the bi-polar jets 
after the initial configuration has been relaxed
(see however e.g. \cite{2020MNRAS.494.3656L} for the recent results with toroidal field initial configurations).
We chose the magnetic field configuration produced by a circular current, same as in \cite{Sap2019}. The only non-vanishing component of the 
vector potential is given by:
\begin{eqnarray}
  A_{\phi}({r,\theta})= A_0 \frac{\left(2-k^2\right) K\left(k^2\right)-2 E\left(k^2\right)}{k\sqrt{4 R r \sin\theta}} \\
  k = \sqrt{\frac{4 R r \sin\theta} { r^2 + R^2 + 2 r R \sin\theta}} \nonumber
\end{eqnarray}
where $E,K$ are the complete elliptic functions and $A_0$ is used to scale the magnetic field and the initial gas to magnetic pressure ratio, $\beta=p_{gas}/p_{mag}$, across the torus.

\begin{figure}
\centering
\includegraphics[width=0.47\textwidth]{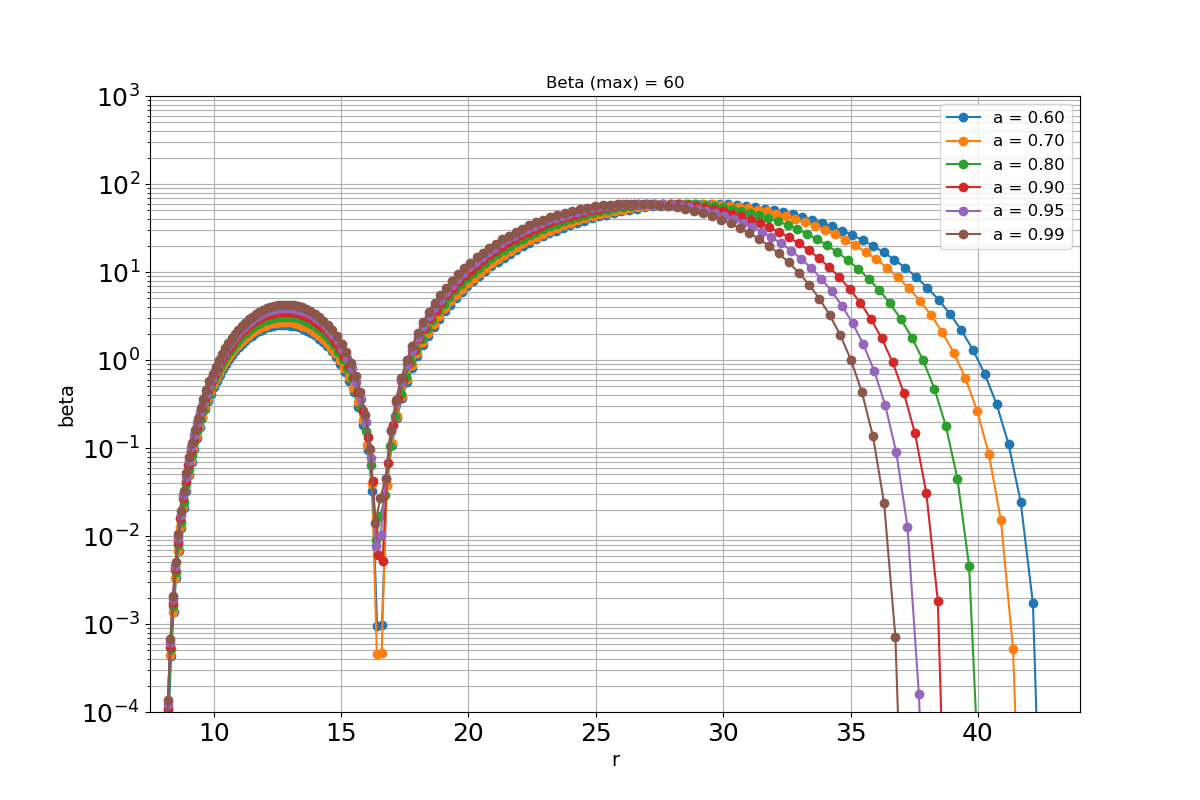}
\includegraphics[width=0.47\textwidth]{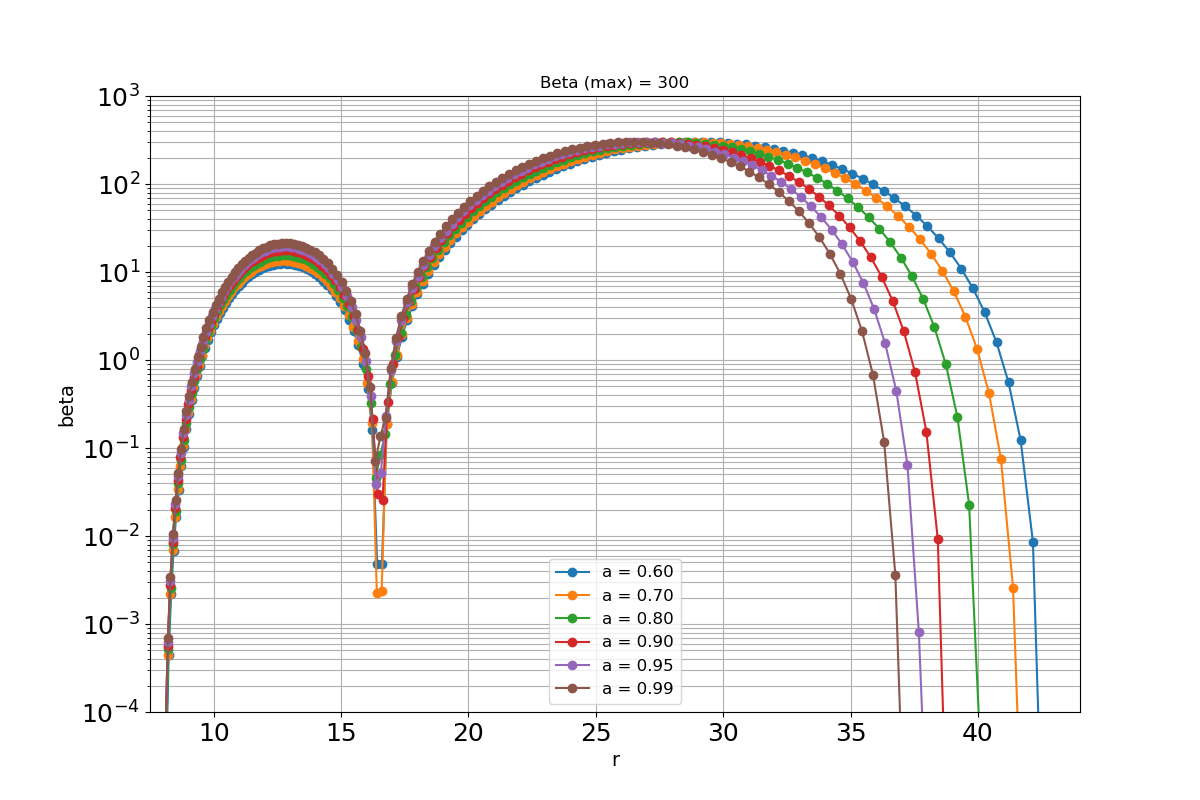}
\includegraphics[width=0.47\textwidth]{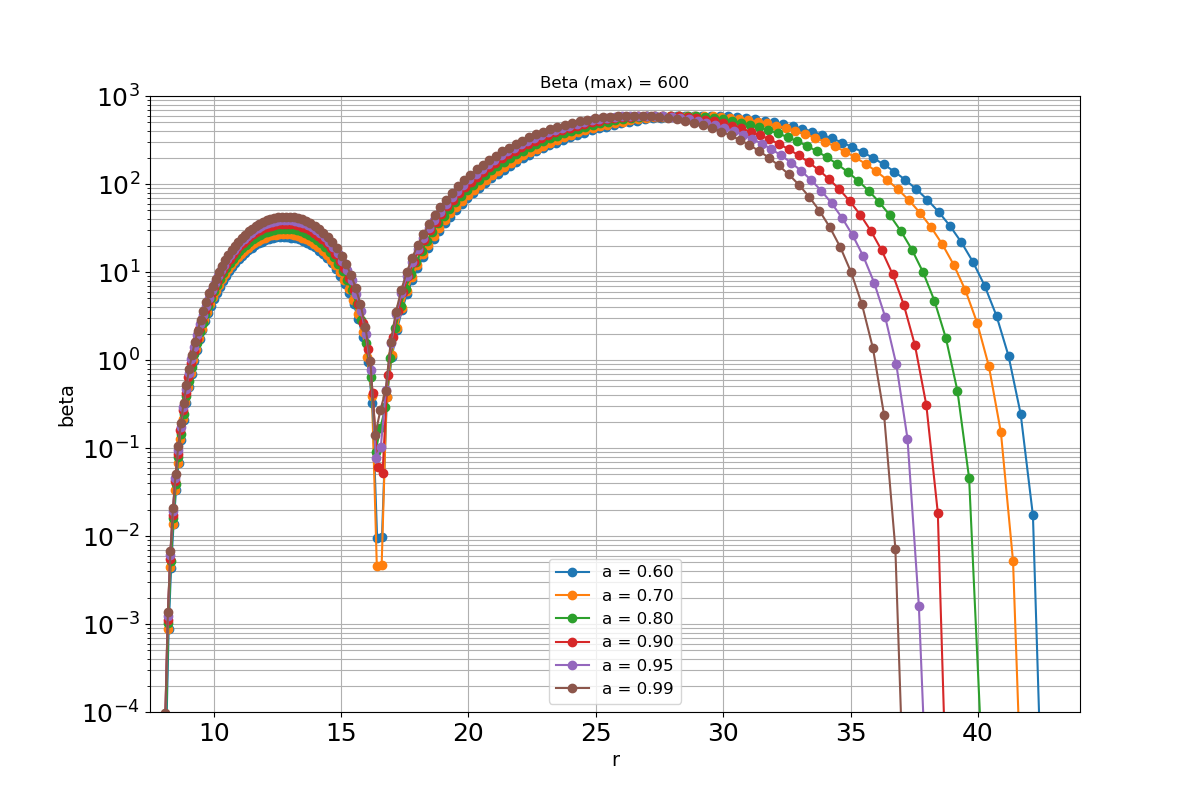}
\caption{Initial distribution of gas to magnetic pressure ratio, $\beta$, in the torus at time t=0.
The models are normalized to $\beta_{\rm max} = 600, 300$, and 60. Values of the Kerr parameter are 
$a=0.6, 0.7, 0.8, 0.9, 0.95, 0.99$.
}
\label{fig_beta_max}
\end{figure}

We define the family of models with varying magnitudes of $\beta$, normalizing them to the maximum value within the torus (the point where $\beta$ is reaching its maximum, depends also on the black hole spin parameter $a$, because of the properties of torus solution for the gas pressure distribution). Our family of models is depicted in Fig. \ref{fig_beta_max}.

In all our simulations we used the resolution of $768 \times 512$ grid points in ($r,\theta$) directions. This allowed us to keep proper MRI resolution, defined as the minimum number of cells per MRI wavelength \citep{siegel}.

\section{Results}
\begin{figure}
\centering
\includegraphics[width=0.47\textwidth]{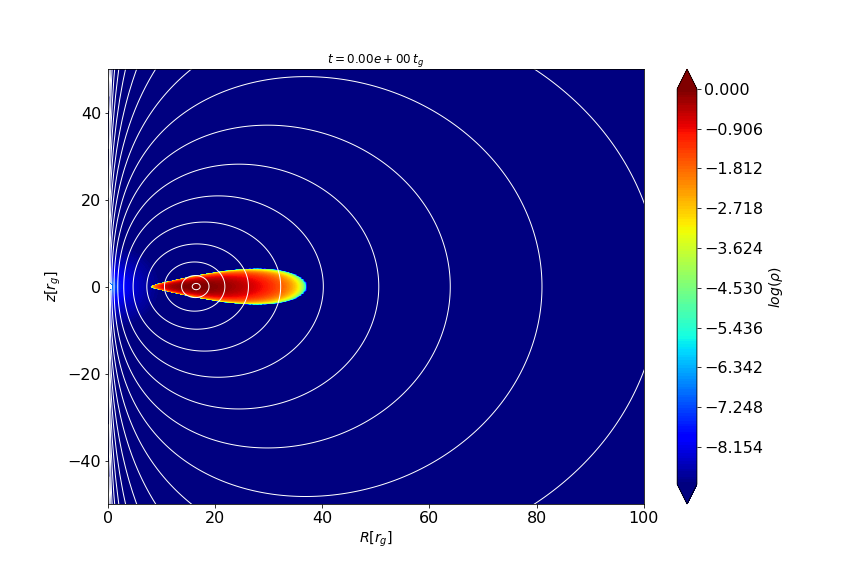}
\includegraphics[width=0.47\textwidth]{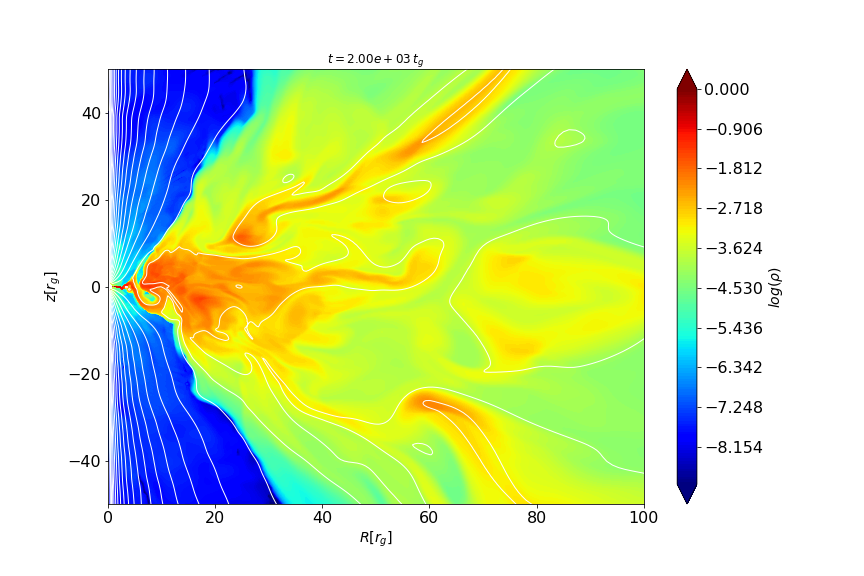}
\caption{Initial conditions and evolved state. Torus density structure and magnetic field contour lines at time t=0 and at time t=2000 M are plotted for the model with $\beta_{max} = 60$ and Kerr parameter $a=0.99$ 
}
\label{fig1}
\end{figure}

The initial configuration of an equilibrium torus as given by the solution of \cite{Chakrabarti1985} is depicted in Fig. \ref{fig1}. Top panel of this figure shows the flattened structure of density enclosed within the region of about $40 r_{g}$. Geometrical thickness of the structure is less than $H/r=0.2$. The solution shown in the plot is parameterized by the black hole spin $a=0.99$.

The inner edge of the torus is $r_{in} = 6 r_{g}$ and the position of maximum pressure in the torus is $r_{max} = 16.5 r_{g}$. All the simulations start from this initial configuration  and evolve afterwards. The value of $\lambda$ ranges between 0.04 to 10.21 M within the torus. The outer radius of the torus is affected by change in Kerr parameter $a$ and $r_{in}^{(von Zeipel)}$.
The surfaces of constant specific angular momentum $l$ and angular velocity $\Omega$ are called the von Zeipel cylinders. With the choice of Chakrabarti's solution for the torus structure, $l= constant$ surfaces become the von Zeipel cylinders. The $r_{in}^{(von Zeipel)}$ parameter represents the true inner edge of the torus and is kept at $8 r_g$ for all the models in the simulation to constrain the outer radius of the torus near to $40 r_g$. 
We note that using bigger values for $r_{in}^{(von Zeipel)}$ would reduce the size of the torus. 
Changing the Kerr parameter from $a = 0.60$ to 0.99 we change the outer radius of torus from 42.5 to $37 r_g$. The average value of $\beta$ within the torus is calculated from the inner point in the torus where the value rises above $10^{-3}$ to the outer point in the torus where the value falls below $10^{-3}$.

The magnetic field imposed on top of the stationary torus configuration has an electric wire shape, with a circular loops concentrated on the radius of pressure maximum. This radius was chosen as $16.5 r_{g}$ for all models.
The magnetic field allows material to start accrete onto black hole. Evolved structure of the flow, which has already relaxed from its initial configuration, is depicted in the bottom panel of Fig. \ref{fig1}.

\begin{figure}
\centering
\includegraphics[width=0.47\textwidth]{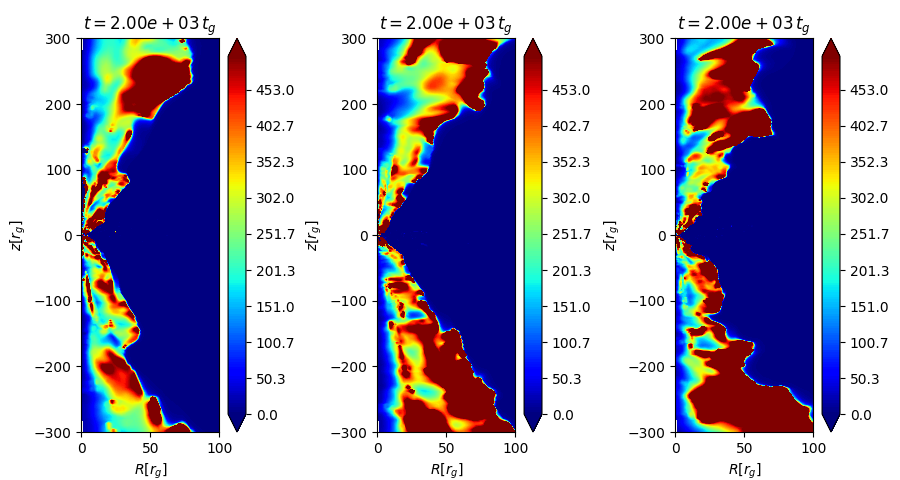}
\includegraphics[width=0.47\textwidth]{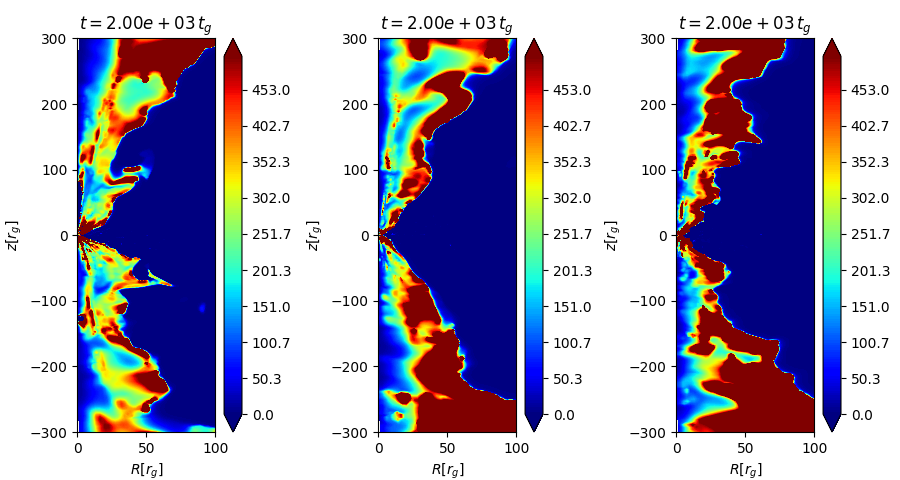}
\caption{Jet structure at time t=2000 M. Plot shows distribution of energetics defined as $\mu$ (see Equation \ref{eq:mu}). 
Top row: models with $\beta_{max}=600$,
bottom row: models with $\beta_{max}=60$.
Models display jets launched from spinning black holes with the Kerr parameter of $a=0.6$ (left column), $a=0.8$ (middle column), and $a=0.95$ (right columns).
}
\label{fig_jets}
\end{figure}

In our simulation the accretion flow has an axisymmetric configuration.
To asses the MAD-ness of the dynamical solution, we would need to cover the non-axisymmetric modes. However, even in our setup, we can evaluate the ratio between the magnetic flux and mass accretion rate on the black hole horizon. The magnetically arrested state appears in fact in the most magnetized models, i.e. those with $\beta_{max}=60$. At the beginning of the simulation, before time $t=2000 ~ t_{g}$, the parameter $\Phi_{BH} = \frac{1}{2\sqrt{\dot{M}}} {\int \mid{B^{r}(r_{H}}) \mid dA_{\theta \phi}}$ is greater than 10 (note, that in our code we use the Gaussian units, so that the factor $4\pi$ is not incorporated in the magnetic flux). 

The jet energetics determines its Lorentz factor at infinity, and as was shown by \citep{Vlahakis2001, Sap2019} is given by
the $\mu$ parameter.
It is defined as: 
\begin{equation}
 \mu = - \frac{T^r_t}{\rho u^r}
\label{eq:mu}
\end{equation}
where  $T^r_t$ is the energy component of the energy-momentum tensor, that consists of gas and magnetic parts, $\rho$ is the gas density, and $u^r$ is the radial velocity, ie. the total plasma energy flux normalized to the mass flux. It is therefore given by the sum of the inertial-thermal energy of the plasma and and its Poynting flux, which can be transferred to the bulk kinetic energy of the jets at large distances.

The distribution of jet energetics parameter in an evolved state of the simulation is shown in Fig. 
\ref{fig_jets}. The snapshots compare two values of magnetic field normalisations, $\beta=600$ and $\beta=60$, in top and bottom rows. We show three different values of black hole spin, $a=0.6, 0.8$, and 0.95. We note a highly inhomogeneous outflows, where larger values of $\mu$ are reached at the edges of the jets rather than at the $z$ polar axis. Already from the color scales of the distribution it can be seen that more energetic jets are produced from fastly spinning black holes, which confirms our intuitions. 
The relation with magnetisation $\beta$ is not that clear though, and seems affected by the black hole spin value.
The details of simulation results are therefore summarized quantitatively in Table \ref{table1}.

The table shows the values of minimum variability timescale and Lorentz factor with the changing black hole spin value. Three models with different magnetic field normalization are shown here. For our calculations the Lorentz factor is taken as the average of $\mu$ in time. The averages were calculated from $t = 600$ to $t = 3100 t_g$. Minimum variability timescale is calculated as the average of peak widths at their half maximum on the $\mu$ variability plot. 

\begin{figure*}
\centering
\includegraphics[width=0.28\textwidth]{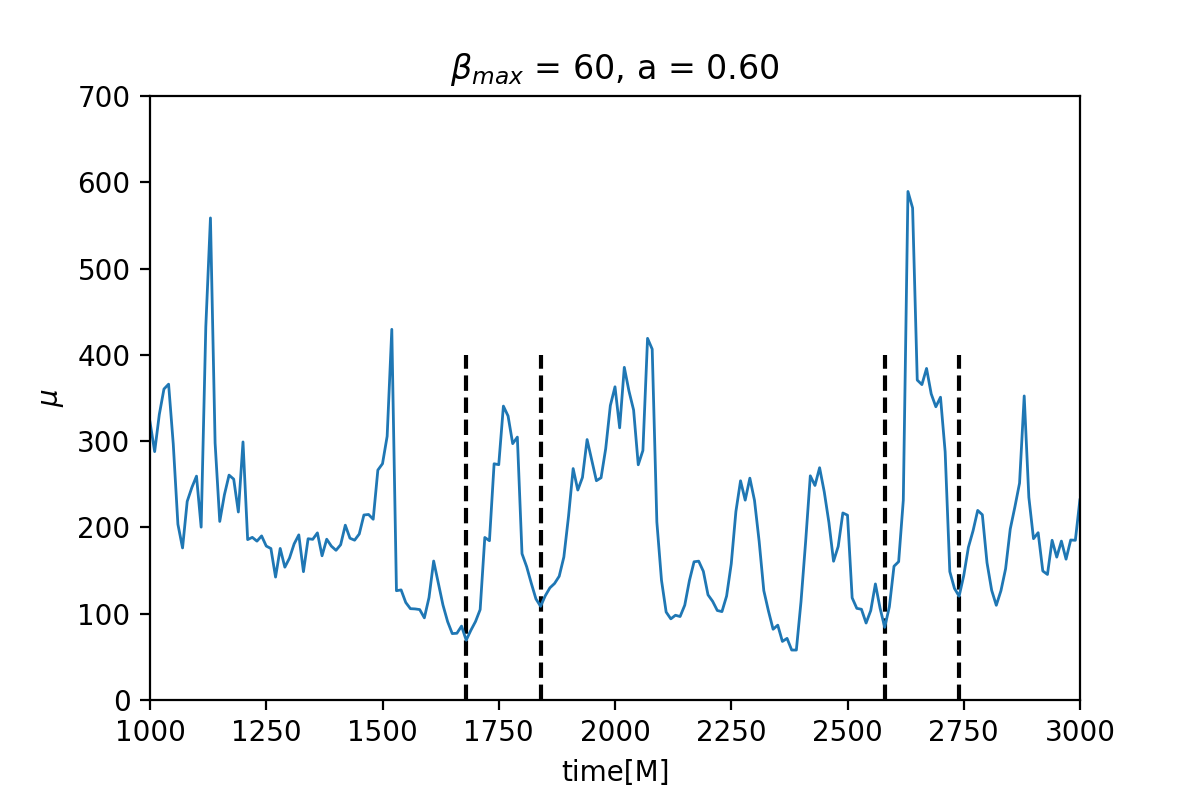}
\includegraphics[width=0.28\textwidth]{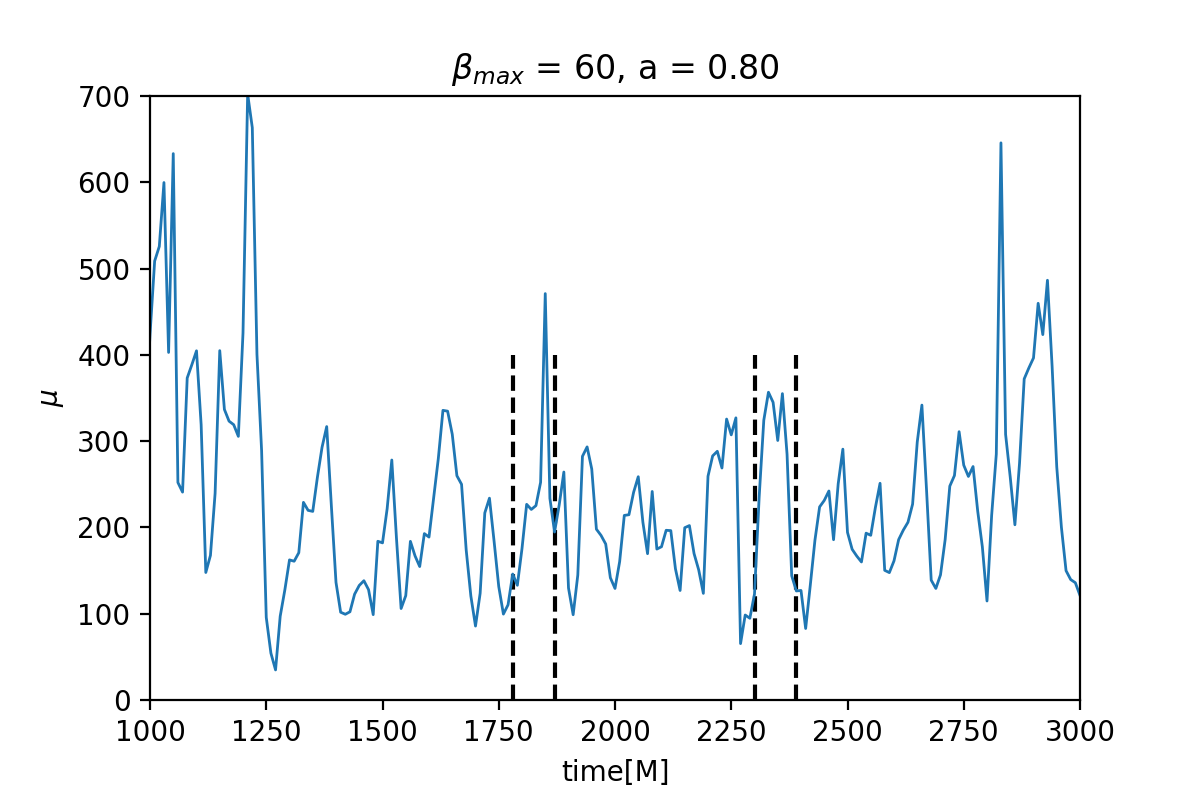}
\includegraphics[width=0.28\textwidth]{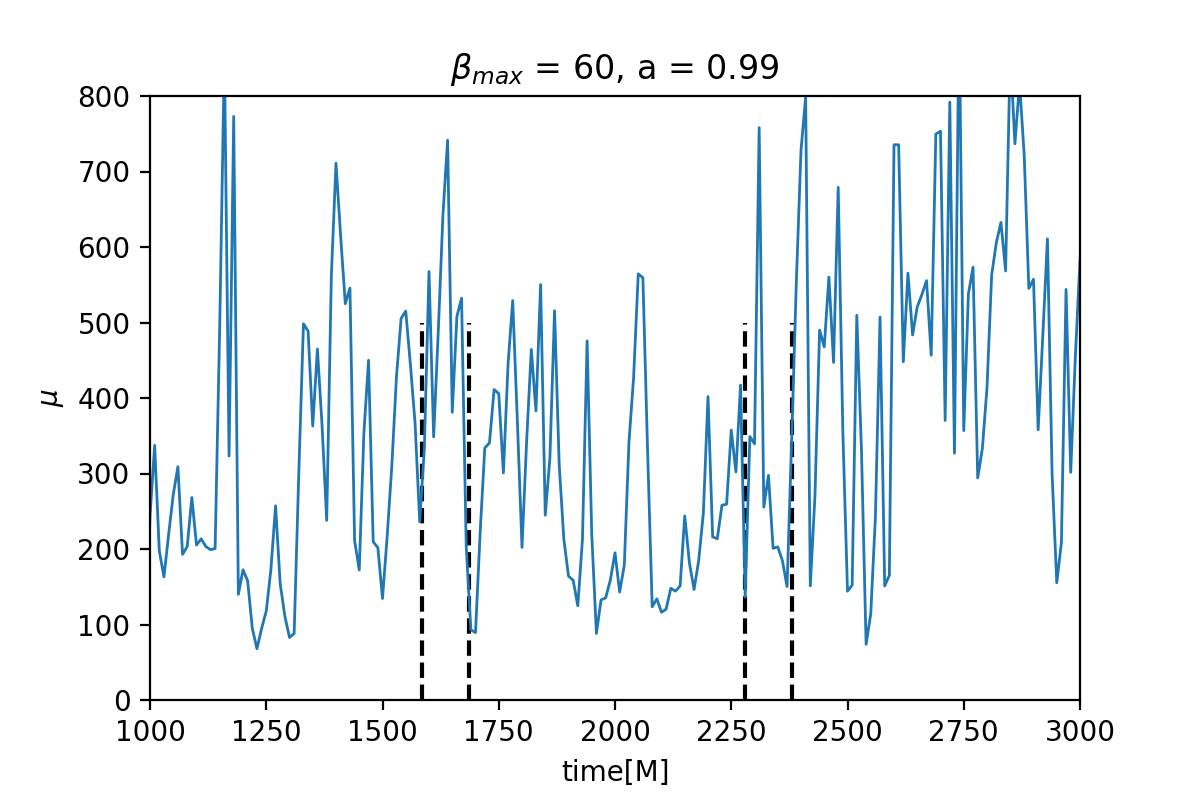}
\includegraphics[width=0.28\textwidth]{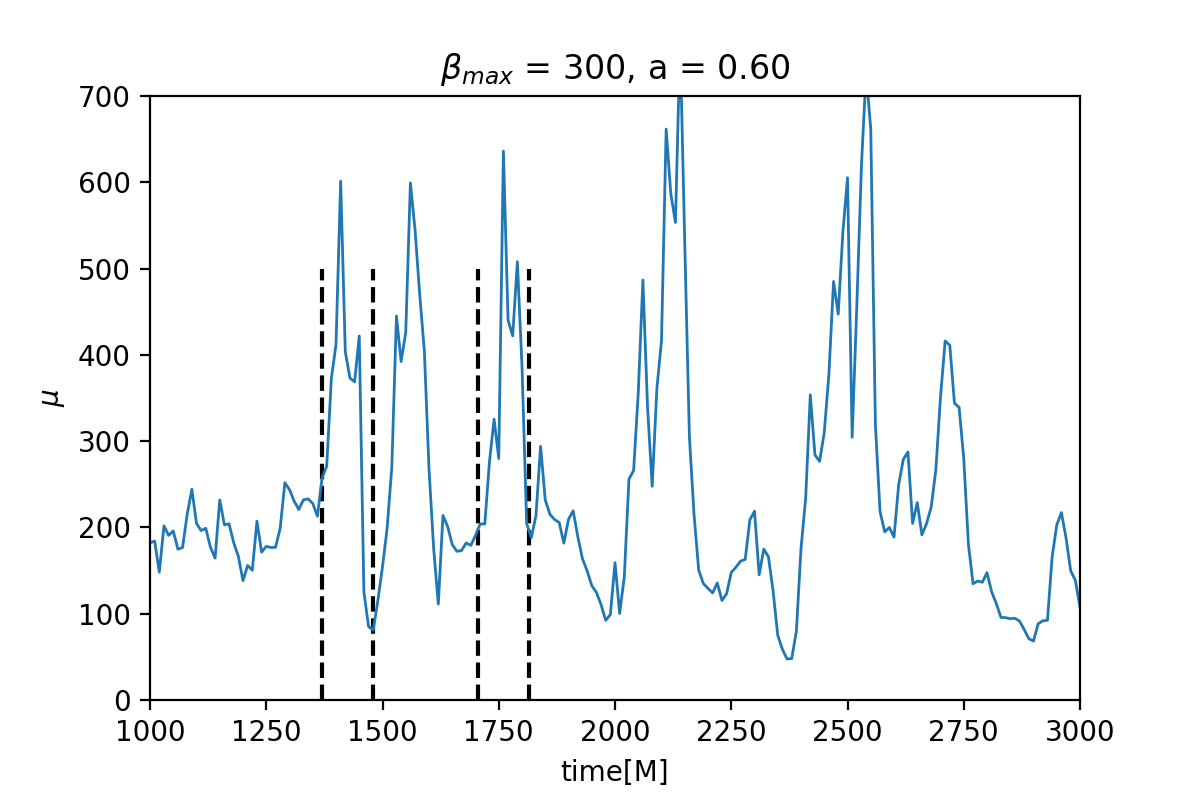}
\includegraphics[width=0.28\textwidth]{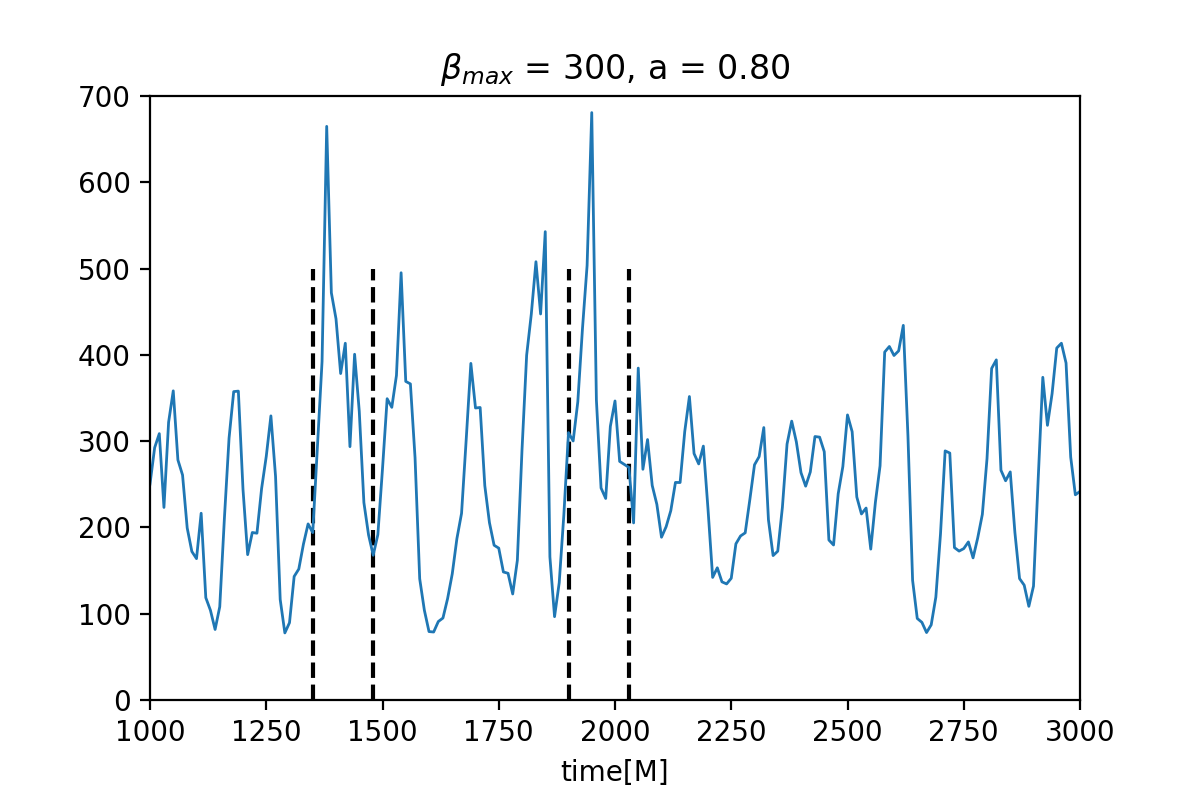}
\includegraphics[width=0.28\textwidth]{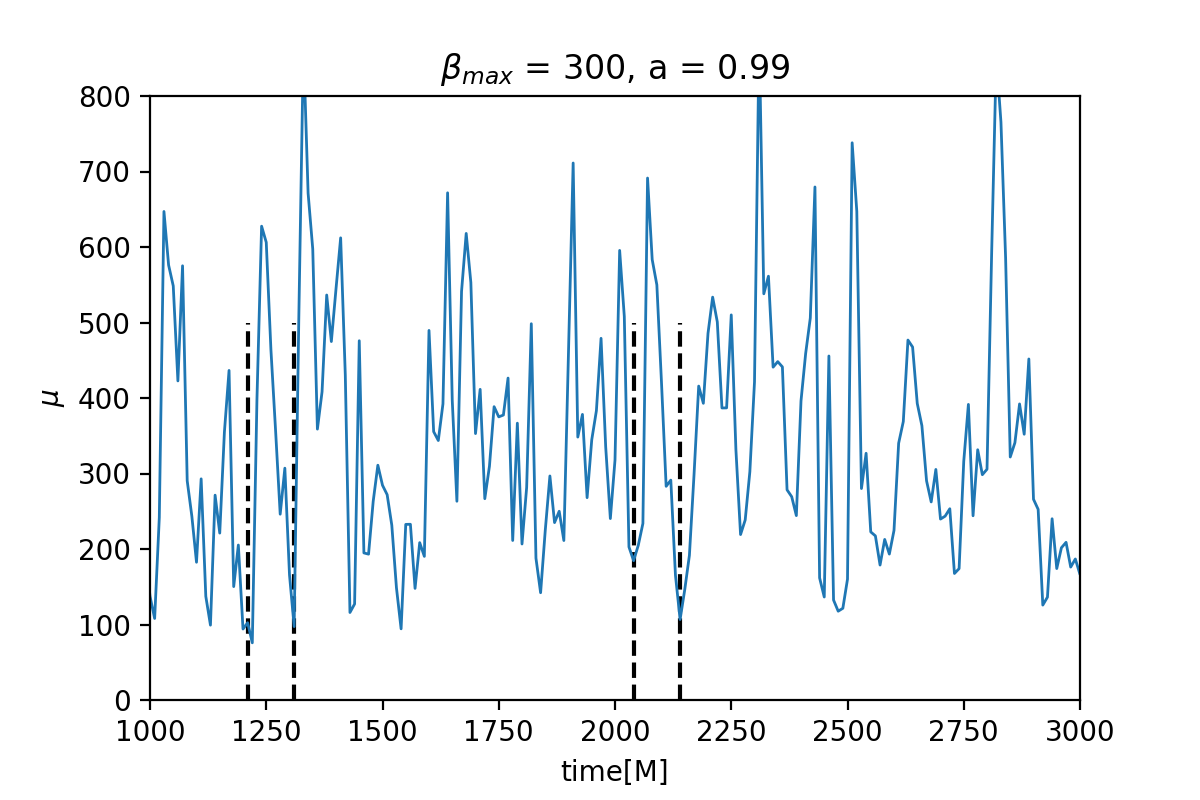}
\includegraphics[width=0.28\textwidth]{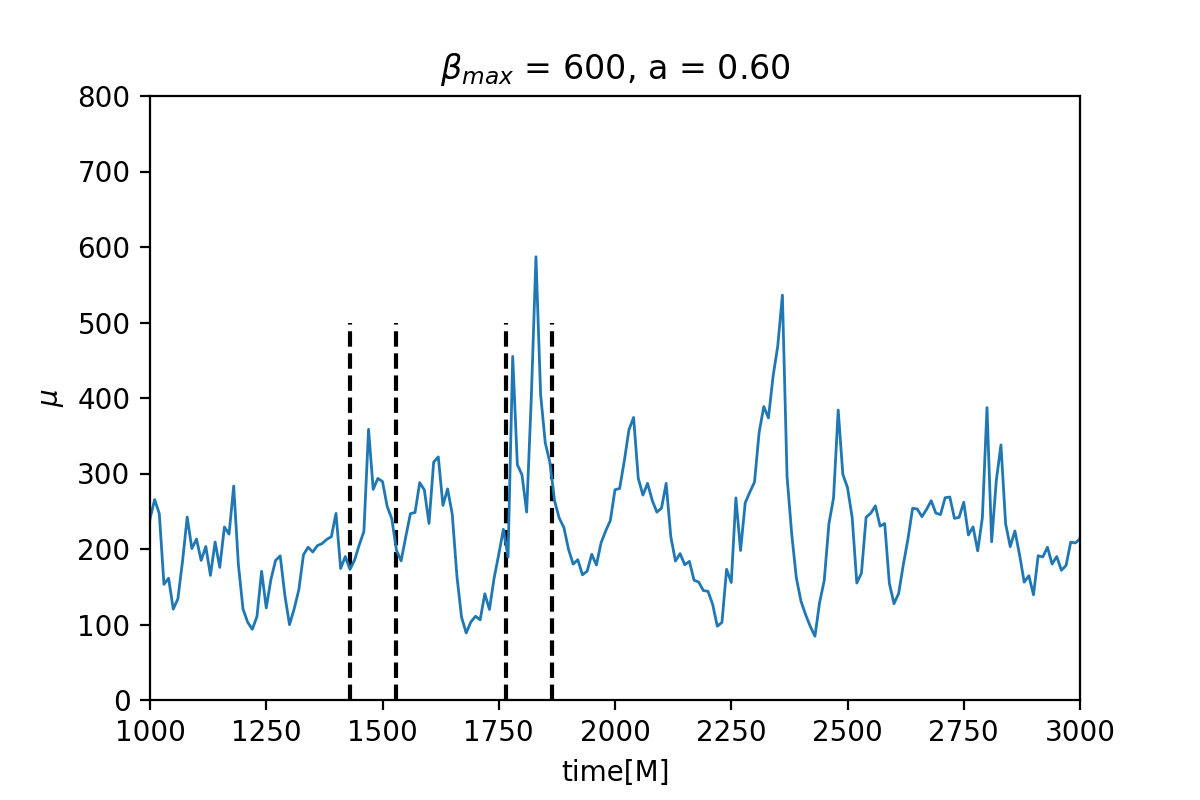}
\includegraphics[width=0.28\textwidth]{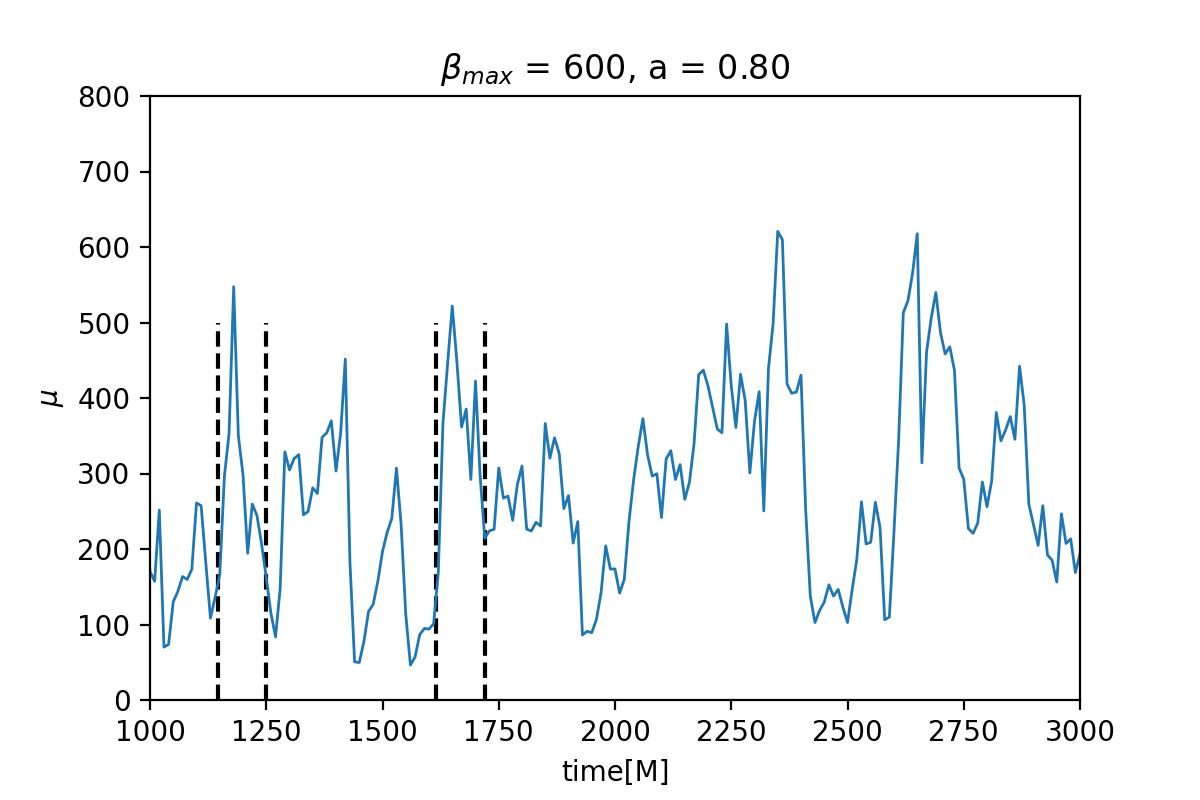}
\includegraphics[width=0.28\textwidth]{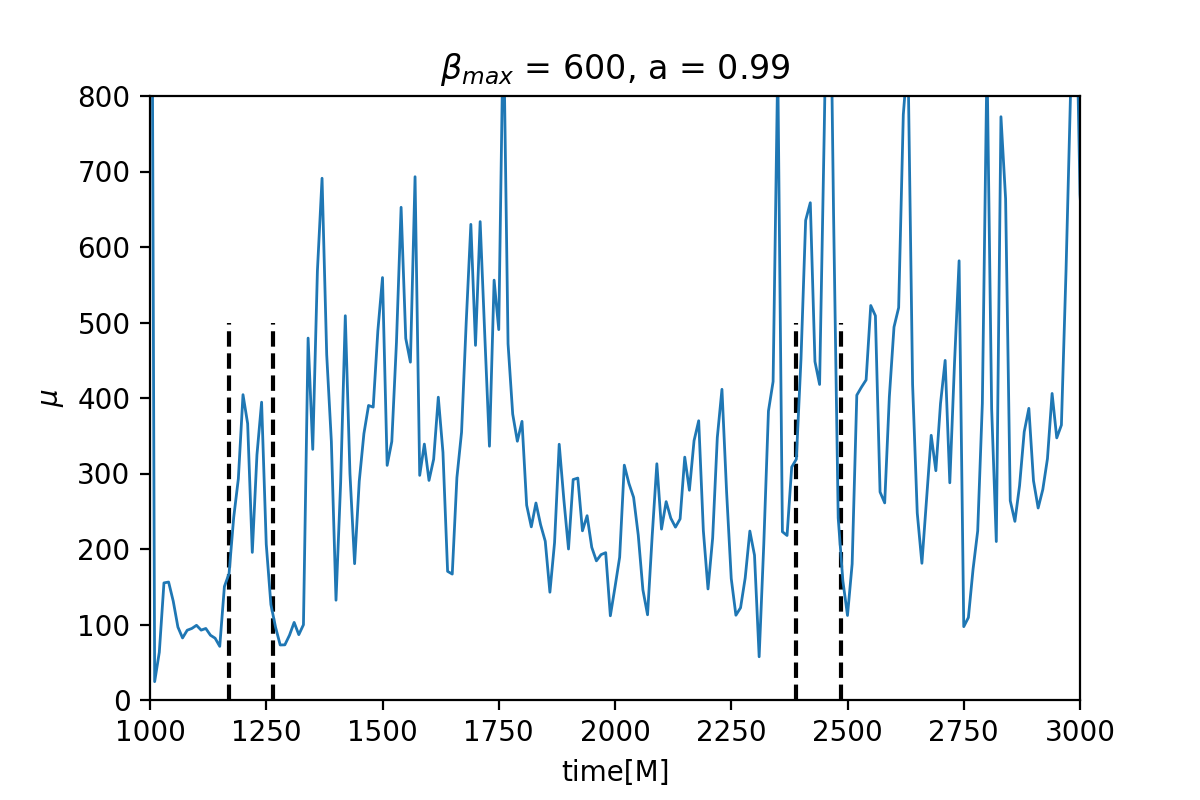}

\caption{Variability of jet  $\mu$-parameter as function of time, for the models with different $\beta_{max}$ (60, 300, and 600, from top to bottom) and three different values of Kerr parameter ($a=0.6, 0.8$, and 0.99, from left to right). The time series are extracted at point p1 in the jet, so at the inclination $\theta=5^{\circ}$ to the vertical axis. The dashed lines represent the characteristic timescale of the MRI calculated using the expression for maximum growth rate derived by \cite{Gammie2004}. }
\label{fig_mu}
\end{figure*}

In Fig. \ref{fig_mu} we show the time variability of jet energetics (i.e. the $\mu$-parameter) for the models with  magnetic field normalization $\beta_{max} = 300$ and different values of black hole spin. 
The variability is measured here at a chosen specific point, $r = 150 r_{g}$ and $\theta = 5^{\circ}$. Here we show the values of $\mu$, from $t = 1000 t_g$ to $t = 3000 t_g$. For each simulation, the parameter $\mu$ is computed at two different points located at $r = 150 r_g$, $\theta = 5 ^{\circ}$ and $\theta = 10^{\circ}$.

\begin{table}[]

	\resizebox{0.48\textwidth}{!}{%
	\hskip-1.65cm
	   \centering
		\begin{tabular}{|c|c|c|c|c|c|c|}
        \hline
        $\beta_{max}$ in torus &  Average $\beta$ & spin (a) & MTS & \multicolumn{3}{|c|}{Lorentz factor $\Gamma$} \\  [0.1cm]
        \hline 
        & & & & Point 1 & Point 2 & Average \\   [0.1cm]
        \hline
			\multirow{6}{5em}{$\beta_{max} = 600$} 
			& 145.54 & 0.99 & 19.73 & 347.26  & 602.29  & 474.77  \\   [0.1cm]
			& 145.86 & 0.95 & 26.40 & 280.66  & 594.80  & 437.73  \\  [0.1cm]
			& 146.87 & 0.90 & 25.34 & 265.89  & 505.47  & 385.68  \\   [0.1cm]
			& 148.43 & 0.80 & 36.88 & 268.89  & 435.86  & 352.38  \\   [0.1cm]
			& 149.70 & 0.70 & 35.56 & 256.27  & 404.21  & 330.12  \\   [0.1cm]
			& 151.01 & 0.60 & 39.66 & 223.22  & 343.72  & 283.47  \\  [0.1cm]
			\hline
			\multirow{6}{5em}{$\beta_{max} = 300$} 
			& 73.35  & 0.99 & 24.20 & 337.80  & 651.93  & 494.87  \\   [0.1cm]
			& 74.07  & 0.95 & 26.55 & 301.88  & 584.01  & 442.94  \\   [0.1cm]
			& 73.99  & 0.90 & 28.31 & 281.92  & 476.51  & 379.21  \\   [0.1cm]
			& 74.76  & 0.80 & 31.98 & 257.70  & 491.84  & 374.77  \\   [0.1cm]
			& 74.85  & 0.70 & 31.61 & 248.67  & 428.22  & 338.44  \\   [0.1cm]
			& 75.50  & 0.60 & 45.65 & 228.90  & 332.53  & 280.71  \\  [0.1cm]
			\hline
			\multirow{6}{5em}{$\beta_{max} = 60$}  
			& 14.91  & 0.99 & 22.91 & 359.43  & 1000.72 & 680.08  \\   [0.1cm]
			& 14.93  & 0.95 & 23.15 & 234.38  & 607.69  & 421.03  \\   [0.1cm]
			& 14.91  & 0.90 & 26.73 & 230.75  & 479.02  & 354.89  \\   [0.1cm]
			& 15.06  & 0.80 & 34.75 & 225.11  & 377.68  & 301.40  \\   [0.1cm]
			& 15.41  & 0.70 & 31.15 & 204.59  & 314.61  & 259.60  \\   [0.1cm]
			& 15.42  & 0.60 & 43.14 & 192.39  & 309.97  & 251.18  \\  [0.1cm]
			\hline
		\end{tabular}
 	}
\caption{Summary of the models studied. Three families of models differ with respect to the magnetic fields normalisation, which are scaled with the maximum value of the gas to magnetic pressure ratio within the torus (note that $\beta(r_{max})$, the value at the radius of pressure maximum, can be as small as $10^{-3}-10^{-2}$, see Fig.\ref{fig_beta_max}, so that all our models are essentially representing strongly magnetized tori). We also give the value of the average value of $\beta$ in the second column. Third column gives the value of the black hole Kerr parameter, $a$, for each model. The resulting variability timescale and Lorentz factor measured as the averaged energetics parameter at two chosen points in the jet, are given in the last two columns.}
\label{table1}
\end{table}

\section{Jet properties and central engine}

We investigate here the influence of the central engine properties, as scaled by its magnetisation, and Kerr parameter $a$ of the black hole, on the variability and energetics of the jet.
The total energetics is described by the parameter $\mu$, which represents the total, thermal and Poynting energy in the jet.
Note that this parameter is dimensionless, as it is given by the ratio of the $r$-component of the linear momentum, to the mass flux across the radial surface (see Eq.~\ref{eq:mu}).
Therefore, it can be
related  to the maximum achievable Lorentz factor, reached at 'infinity', and available under the 'infinite' efficiency of conversion to the bulk kinetic
energy of particles injected to the jet. We identify therefore the time-averaged value of $\mu$ as the proxy of jet Lorentz factor, $\Gamma$. 
The variability is also measured by $\mu$ changes with time, at a given point.
We propose, that the frequency of these changes, measured in the base of the
jet,  is related to the frequency of collisions between the shells transported downstream the jet and being the source of observable gamma-ray pulses, produced in the internal shock scenario 
\citep{Kobayashi1997}.

The jet structure is clearly non-uniform, and more energetic blobs are always located in the outer regions, while less energetic ones travel close to the axis.
This is revealed by the systematic differences between $\Gamma$ measured at point $p1$, which is ranging between $\sim 200$ and 350, and those measured at point $p2$, which is ranging from $\sim 300$ up to 1000 (see Table \ref{table1}, and bottom panel in Figure \ref{correlations}). The jet bulk velocity, and hence its power, increases with black hole spin, and reaches dramatically average values if the black hole rotates close to the Kerr limit. This is expected to be a result of the Blandford-Znajek driven process.
The dependence on the magnetic pressure in the disk, and $\beta$-parameter, is however not linear.
Only in the case of the most spinning black hole, $a=0.99$, the most magnetized disk, with average $\beta$ in the torus on the order of 15 ($\beta(r_{max})<10^{-3}$ and $\beta_{max}=60$), the jet power is largest, than for more thermal ensure dominated tori.
If the black hole does not rotate at close to the maximum Kerr limit, then
the more thermally dominated tori, with average $\beta_{max}$ of 300, or 600,
(average $\beta$ in the torus is 75 or 150, respectively) give more power to the jet. This result can be understood in the frame of magnetically driven
transfer of accretion disk energy to the jet, when there is less Poynting flux available in the funnel for less spinning black holes, while the thermal
energy can still be transported through the horizon with enough efficiency.
Notably, there is almost no difference between the jet power and
angle-averaged Lorentz factors, in the $\beta_{max}= 300$, and 600 cases. For spins $0.7\le a \le 0.9$ the point $p2$ meets more energetic blobs for smaller $\beta$, while blobs at point $p1$ are found more energetic for lager $\beta$.

\begin{figure}
	\centering
	\includegraphics[width=0.47\textwidth]{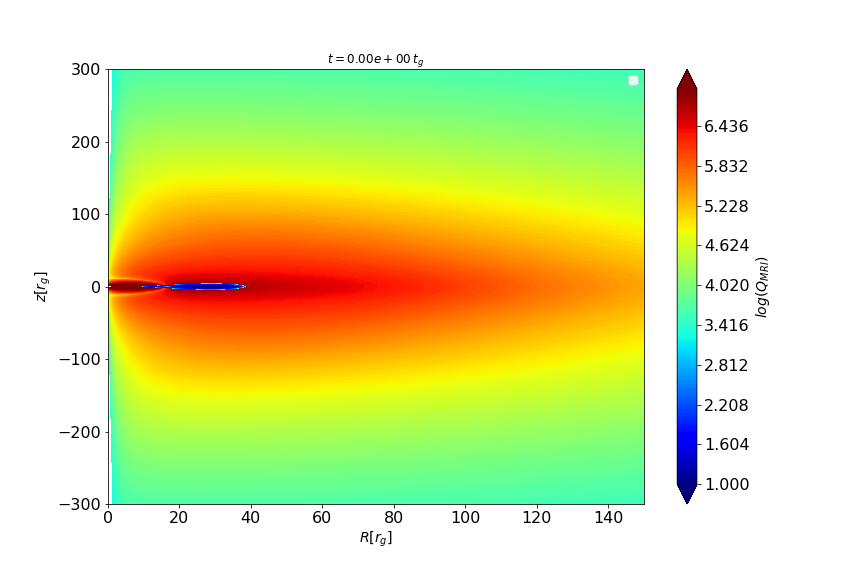}
	\includegraphics[width=0.47\textwidth]{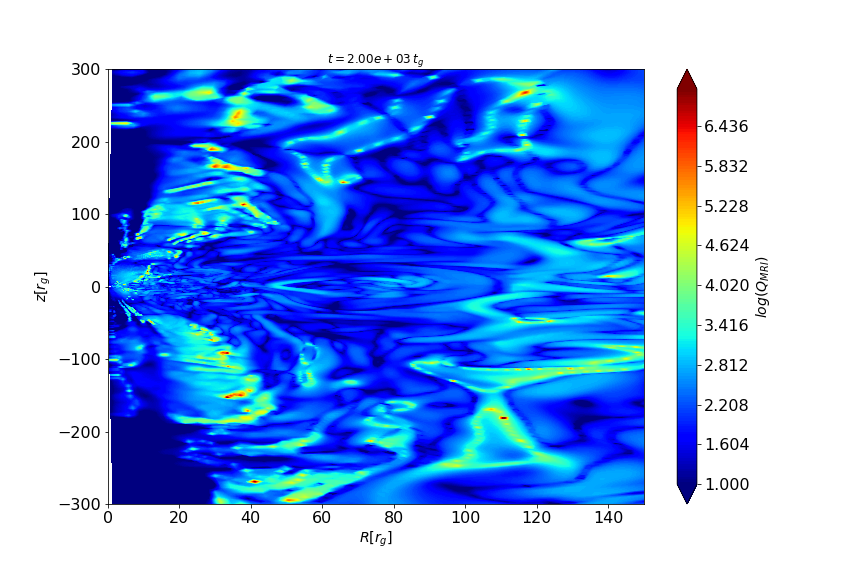}
	\caption{Color contour map of $Q_{MRI}$, defined as the number of grid cells per the MRI wavelength, shows in the logarithmic scale at time t=0 and at time t=2000 M. Plots are made for the model with $\beta_{max)}$ = 300 and Kerr parameter $a=0.9$ 
	}
	\label{qmri}
\end{figure}

\begin{figure}
	\centering
	\includegraphics[width=0.5\textwidth]{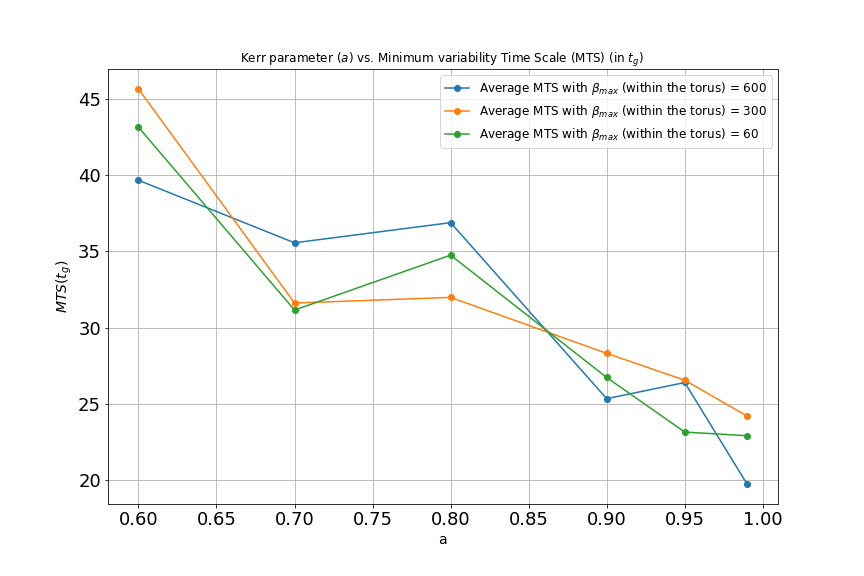}
	\includegraphics[width=0.5\textwidth]{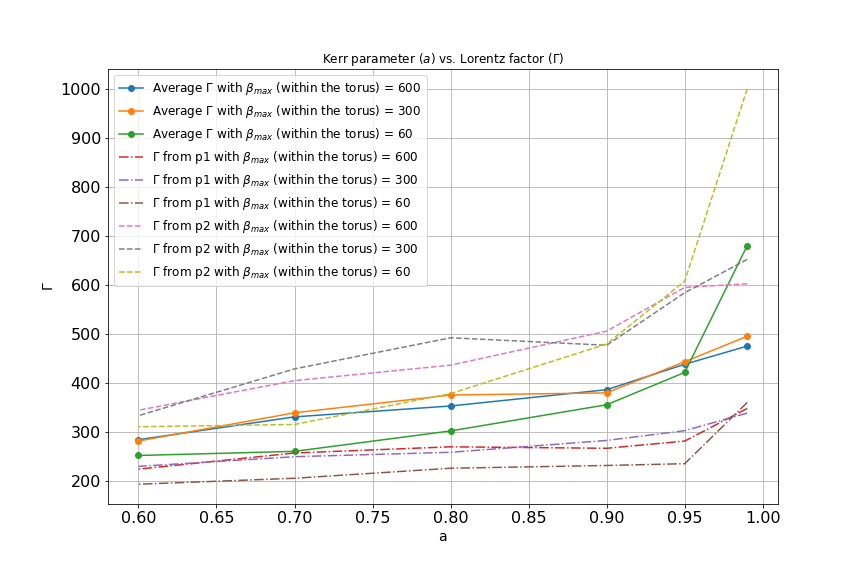}
	\caption{Correlations between (a) the Kerr parameter $a$ and Minimum variability Time Scale and (b) the Kerr parameter $a$ and Lorentz factor. The upper panel shows results for three families of models, differing with average (and maximum) $\beta$ parameter: $\beta_{max}=60$ (green), 300 (orange) and 600 (blue). The MTS timescale is computed as the average duration of the pulses in the $\mu$ time series. The values given are averages from the two points p1 and p2. 
          The bottom panel shows results for the $\Gamma$ factor, defined as the
          average energetic parameter $\mu$, measured from time 600 till 3000 $t_{g}$. Long-dashed lines represent measurements at point p1 in the jet, short-dashed are for point p2, and solid lines are the  average between two points. 
	}
	\label{correlations}
  \end{figure}

The variability of the jet, studied in terms of the pulses duration, is driven by the magneto-rotational instability in the disk.
  Here however also some numerical constraints of our simulation, namely spatial resolution, and the axisymmetric setup of the models, may be of some importance. The MRI is resolved in terms of the minimum number of cells per wavelength, as shown in Figure \ref{qmri}. Nevertheless, the pulses duration only roughly correlate wit $t_{MRI}$ (Fig. \ref{fig_mu}), and only the widest pulses are clearly showing this effect. The narrower pulses, which contribute also to our minimum variability timescale, MTS estimate, behave more erratically.
    Therefore, as displayed in Fig. \ref{correlations} (upper panel), the MTS
  has a general trend to decrease with the black hole spin, but it can either decrease o increase with $\beta$, depending on the $a$ value. In particular, we can note that the thermal pulses are shortest for $a=0.95$ while they are longest for $a=0.8$.
  
\begin{figure}
	\centering
	\includegraphics[width=0.5\textwidth]{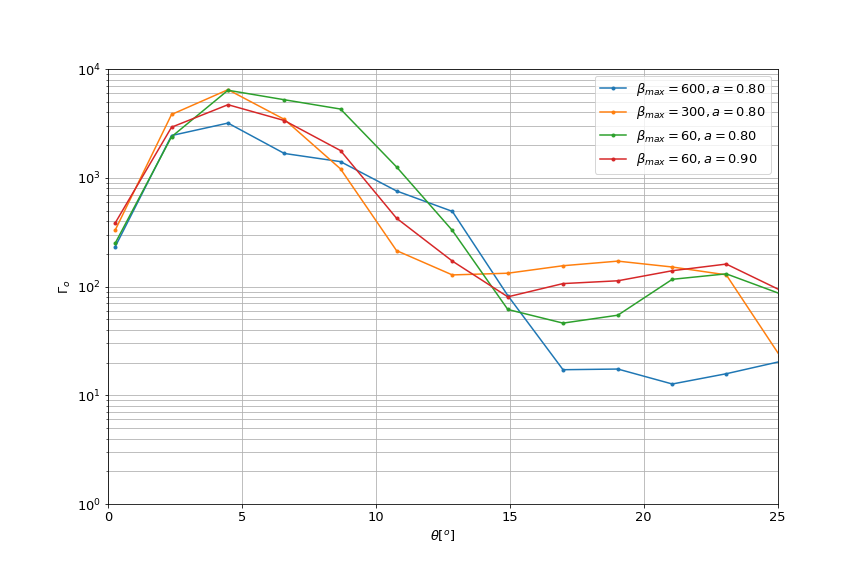}
	\caption{The time-averaged jet Lorentz factor as measured at a distance of 2000 $r_{g}$, in the function of polar angle $\theta$. The plot shows results for four chosen models, differing with maximum $\beta$ parameter: $\beta_{max}=60$ (red and green), 300 (orange) and 600 (blue). The black hole spin was either $a=0.8$ or $a=0.9$, as marked in the plot.
	}
	\label{fig_jetprofile}
  \end{figure}

We note that important information about the jet engine and jet collimation comes from the angular jet structure. Our jet is not uniform and has a distribution of energy content that is both time- and angle dependent. We probed here how the jet distributes its power and we plot $\Gamma$ as function of polar angle. We calculated the time-averaged "jet profile" at a radius of 2000 ~$r_{g}$, so at a large distance from the black hole. It is presented in Fig. \ref{fig_jetprofile}.
  The profile shows that most energetic part of the jet is located inside a narrow region $\theta<15^{\circ}$ which is qualitatively very similar to the profiles found in recent 3D black hole jet studies (\citep{Kathir2019}, see also \cite{Nathanail2021}). Compared with those results, our jets are accelerated to a larger $\Gamma$, for the same black holes spin. This is due to our magnetisation profile and initial different $\beta$ distribution in the torus, which in those works has been adopted uniform and larger on average \citep{Fernandez2019}.

Finally, we verified if our results depend on the adopted value of the density floor, i.e. the numerical floor in our simulation. The minimum density in our runs is forced to never drop below $\rho_{min}=10^{-7}$. As shown already in \cite{Sap2019}, the time-averaged value of the energetics $\mu$-parameter converged for various adopted density floors (see Figure 5. in their paper).
Similarly here, even though we are using different initial conditions and magnetisation in the current models, the density floor value does not significantly affect the time averaged results, provided it is sufficiently low. We show our testing results in Table \ref{densityfloors}, where we compare the time averaged Lorentz factors at the two distinct points in the jet. The testing model was used with parameters $a=0.9$ and $\beta_{max}=60$. 

We also checked that the variability minimum timescale, MTS, calculated at different locations in the jet depends somewhat on the density floor value, but the results do not follow any specific trend. In general, MTS values at point p2 are always smaller than in point p1, and their ratio is about 2/3 (with an exception of the floor 1.e-9, where the ratio is almost 1/2).

\begin{table}[h]
\begin{tabular}{|l|l|l|l|}
\hline
                               & \multicolumn{2}{c|}{Time-averaged $\Gamma$} \\ \hline
Density floor & From point p1  & From point p2 \\ \hline
1.e-17                         & 258.28         & 469.20     \\ \hline
1.e-15                         & 267.20         & 577.17      \\ \hline
1.e-12                         & 250.00         & 421.46        \\ \hline
1.e-9                          & 237.84         & 472.90       \\ \hline
1.e-7 (original simulation)   & 230.75         & 479.02        \\ \hline
1.e-5                          & 113.15         & 239.35        \\ \hline
\end{tabular}
\caption{The dependence of jet Lorentz factor (computed from $\mu$-parameter) on the adopted density floor}
\label{densityfloors}
\end{table}

  In order to better understand the jet variability in our models,
  and also to be able to compare it to observed light curves originating from gamma ray emission at large distances, we performed the time series analysis of our modeled sequences.

\subsection{Time series analysis}

\begin{figure}
	\centering
	\includegraphics[width=0.5\textwidth]{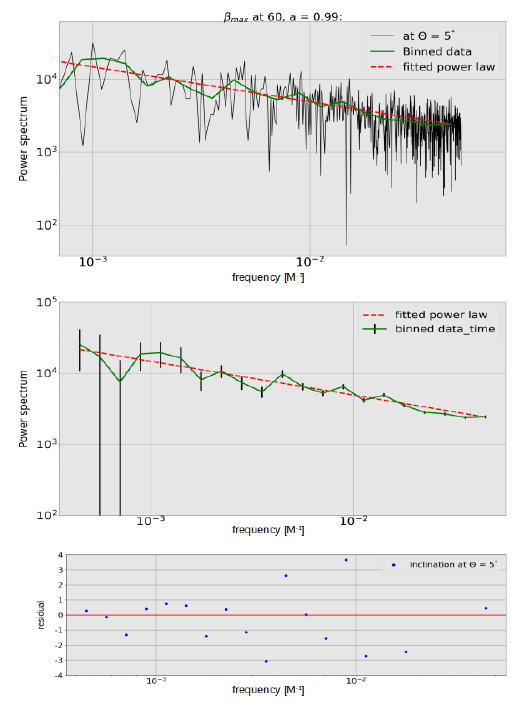}
	\label{PDS}
	\caption{The top panel shows the simulated time series along with the binned data and the fitted power law. Middle panel shows the error bar in the binned data and the bottom panel is the residual plot for binned data and fitted PL (Power Law). The plots in this figure correspond to the model with $\beta_{max}=60$ and spin $a=0.99$ at an inclination of $\theta=5^{\circ}$. }
  \end{figure}

We consider the time series of the $\mu$-parameter (defined in Eq.~\ref{eq:mu}) in order to do the Fourier and Power Density Spectral (PDS) analysis of it. We further impose the logarithmic binning to this time series and we plot the averaged values over the bins. Fig. \ref{PDS} shows our simulated data, in logarithmic scale, corresponding to the model with $\beta_{max} = 60$ and spin $a=0.99$. The jet variability is extracted at an inclination angle of $\theta=5^{\circ}$. The plot shows binned data with PL fitting. The error bars can be seen very large at low frequencies in the panel 2 of Fig. \ref{PDS} as the data are largely spread around the mean value but gradually they decrease, giving perfectly binned data. We further fit the binned data with a Power Law function (PL) - $y(x) = A x^{\alpha}$.
Panel 3 of Fig. \ref{PDS} shows the residuals in the function of frequency and as it can be seen, the fitting is better in the low frequency range. Out of our 18 models, we choose the PDS plot for this model (Fig. \ref{PDS}) because its chi-square value is the lowest among all other models (reduced $\chi^{2} = 14.21$). We note, that in fact the power-law model might not be the best to reproduce the jet variability in this model. On the other hand, there are no significant peaks at specific frequencies, which would be found in the PDS analysis. Also, our aim is to show the general trends and correlations between the central engine properties, and variability probes in the modeled jets. Therefore, we limit our study below to the relations between 

\subsection{Relation between central engine and jet variability}

\begin{figure}
	\centering
	\includegraphics[width=0.5\textwidth]{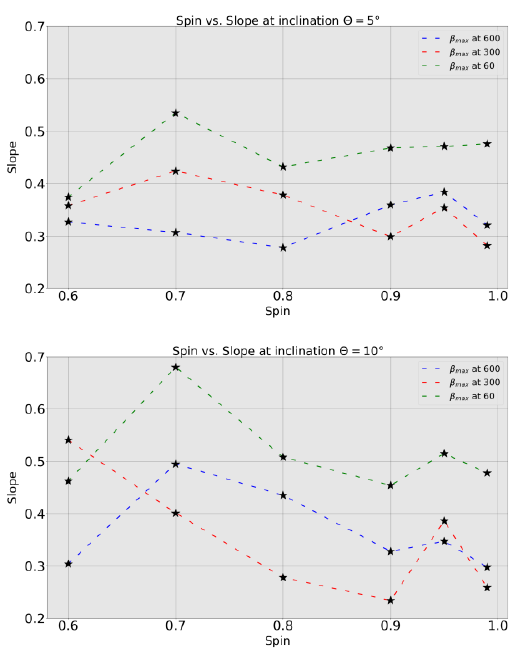}
	\label{slope}
	\caption{Relation between the black hole spin and the power law slope of the fitted PDS for all models, $\beta_{max}=600$ (blue lines), $\beta_{max}=300$ (red lines),  and $\beta_{max}=60$ (green lines). The first panel is chosen at an inclination of $5^{\circ}$ and the second one is at $10^{\circ}$.  }
  \end{figure}

In Figure \ref{slope} we show the relation between the slope of the power aw function and the black hole spin in the jet engine. The time variability of jet energetics (the $\mu$-parameter) is measured at two different inclination angles, $\theta=5^{\circ}$ and $\theta=10^{\circ}$. We show here three families of models, with different magnetisation of the torus. The values of the PL slope fitted for all models with different spins are listed in Table \ref{table2}.

It can be inferred from our analysis that model with lowest torus magnetization, i.e $\beta_{max} = 600$, has steepest PDS, and the highest slope of the power law function is found at spin $a=0.7$. This is measured at both inclinations chosen for observing the jet variability. The model with higher magnetization, i.e $\beta_{max} = 60$ is found to have steeper PDS slopes at higher spin, $a=0.95$,  when measured at an inclination of $\theta=5^{\circ}$. As the inclination increases to $\theta=10^{\circ}$, PDS is steeper at spin $a=0.7$. For an intermediate model with $\beta_{max} = 300$, PDS is always steeper at lower spins, $a= 0.7$, and $a=0.6$, for inclinations of $\theta=5^{\circ}$ and $\theta=10^{\circ}$, correspondingly.

\begin{table}[]
\centering
\begin{tabular}{ |c|c|c|c|c| } 
\hline
 $\beta_{max}$ in torus &  spin $a$ & \multicolumn{2}{|c|}{Slope} \\
\hline & &Point 1, $\theta=5^{\circ}$ &Point 2, $\theta=10^{\circ}$ \\
\hline
\multirow{3}{5em}{$\beta_{max} = 600$ } 
&0.99 &0.321 $\pm$ 0.053 & 0.298 $\pm$ 0.052 \\
&0.95 &0.384 $\pm$ 0.068 & 0.347 $\pm$ 0.056 \\
&0.90 &0.359 $\pm$ 0.081 & 0.328 $\pm$ 0.053 \\
&0.80 &0.278 $\pm$ 0.066 & 0.435 $\pm$ 0.091 \\
&0.70 & 0.308 $\pm$ 0.063 & 0.4495 $\pm$ 0.051 \\
& 0.60 & 0.327 $\pm$ 0.078 & 0.304 $\pm$ 0.052 \\ 
\hline
\multirow{3}{5em}{$\beta_{max} = 300$ } 
&0.99 & 0.282 $\pm$ 0.068 & 0.259 $\pm$ 0.038 \\
&0.95 & 0.354 $\pm$ 0.067 & 0.386 $\pm$ 0.053 \\
&0.90 & 0.298 $\pm$  0.051 & 0.234 $\pm$ 0.071 \\
&0.80 & 0.378 $\pm$  0.063 & 0.278 $\pm$ 0.077 \\
&0.70 & 0.423 $\pm$ 0.092 & 0.401 $\pm$ 0.064 \\
&0.60 & 0.358 $\pm$ 0.063 & 0.542 $\pm$ 0.162 \\
\hline
\multirow{3}{5em}{$\beta_{max} = 60$ } 
&0.99 & 0.476 $\pm$ 0.073 & 0.478 $\pm$ 0.067 \\
&0.95 & 0.471 $\pm$ 0.095 & 0.515 $\pm$ 0.047 \\
&0.90 & 0.468 $\pm$ 0.101 & 0.454 $\pm$ 0.067\\
&0.80 & 0.432 $\pm$ 0.049 & 0.508 $\pm$ 0.049 \\
&0.70 & 0.535 $\pm$ 0.074 & 0.680 $\pm$ 0.123 \\
&0.60 & 0.374 $\pm$ 0.094 & 0.462 $\pm$ 0.121 \\
\hline
\end{tabular}
\caption{Slopes of the fitted power law for all the three models $\beta_{max} = 600$, $\beta_{max} = 300$ and $\beta_{max} = 60$.}
\label{table2}
\end{table}

\begin{figure}
	\centering
	\includegraphics[width=0.5\textwidth]{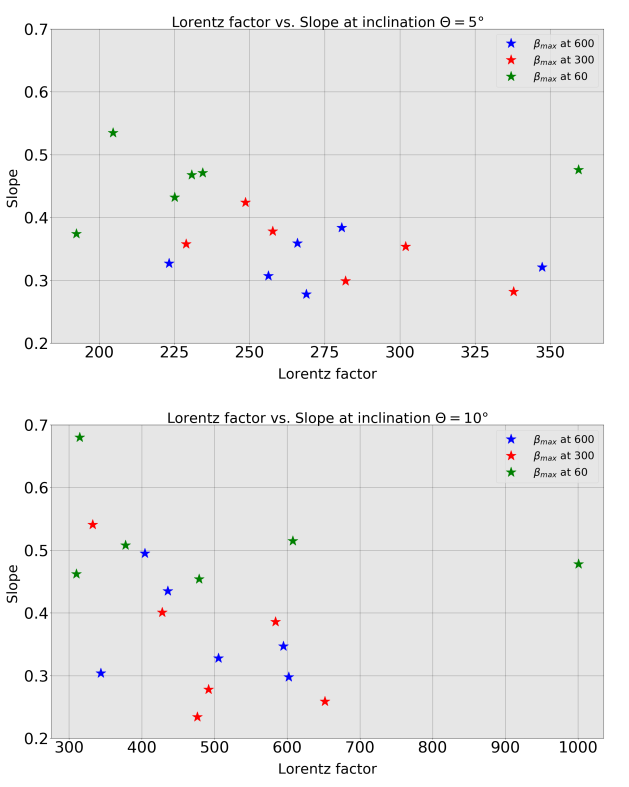}
	\label{fig_stars}
	\caption{Relation between the Lorentz factor in the remote jet and the slope of the fitted power Law for all models, $\beta_{max} = 600$ (blue stars), $\beta_{max} = 300$ (red stars) and $\beta_{max} = 60$ (green stars). The first panel is for the observers inclination of $\theta=5^{\circ}$ and the second one is at $\theta=10^{\circ}$.  }
  \end{figure}

We also investigated how the slopes of the PDS behave for different Lorentz factors of the jets in different models 
(see Figure \ref{fig_stars}). For particular magnetisation values of $\beta_{max}$, we do not see any particular pattern. The PSD are steep, with slopes $(\alpha) \geq 0.55$ for model with lower magnetization i.e $\beta_{max} = 600$ only (see Tables \ref{table1} and \ref{table2}). 
Other models have quite flat PDS spectra, with slopes $(\alpha) \leq 0.55$. The model with $\beta_{max} = 300$ presents the most varying relation between the slope vs. Lorentz factor, as measured  at inclination $\theta=10^{\circ}$.

On the other hand, a general anti-correlation between the jet Lorentz factor and PDS slope of the power-law fit is seen, when we abandon the dependence on the central engine magnetisation. In other words, if the particular GRBs are treated individually, then most of them follow the trend of decreasing PL slope with increasing speed of the jet.
The only outlier from this trend is the GRB which represents the model of most highly magnetized torus around the fastest spinning black hole. Here the jet is very strong, the Lorentz factor reaches even $\Gamma=1000$ in some parts of the jet, while the slope of its variability PL function is again steeper.
  
\section{Discussion and conclusions}

The variability of emission observed in the gamma ray bursts is a complex phenomenon. From the observational point of view, the detected gamma ray flux exhibits a large variety of patterns that reflect complicated processes governing the high energy radiation \citep{Fishman94}. The flux varies on multiple timescales, and power spectral density of the lightcurves is frequently fitted with the power-law function ($P(f)\sim f^{-\alpha}$). The values of slope fitted to individual PDS spectra have a wide range. For the stochastic process driven by internal turbulence in the jet interior, a slope of $\alpha=5/3$ is theoretically expected within the internal shock scenario \citep{Beloborodov2000}. Also, \citep{Zhang2014} proposed a turbulence scenario with magnetic reconnections in the ejected shells to explain a power law PDS shape of the Swift GRBs.

In some GRBs the quasi periodic oscillations have been tentatively detected with a periodicity between 2 and 8 seconds for long and a few milliseconds for short events. These
oscillations can be attributed to the non-steady accretion in the central engine of collapsing star \cite{Masada2007} or to the modulation caused by the black hole spin misalignment, after the merger with neutron star \citep{Stone2013}.

Furthermore, several correlations between the variability properties and GRB energetics have been detected. The peak energy is anti-correlated with the PDS index \citep{Dichiara2016}.
The general idea behind the correlations  this kind invokes the jet Lorentz factor, $\Gamma$, being the main driver
responsible relations between both peak energy an luminosity, and GRB duration  and its luminosity \citep{Dainotti2017}.
The duration of the burst, $T_{90}$,
was also found to be related with the minimum variability timescale (MTS). In the sample of long and short duration GRBs detected by Fermi, the statistical significance for the bi-modal distribution of the events is higher, when the MTS is taken into account
\citep{Tarnopolski2015}.

In our simulations, the variability in the jet is related to the action of the central engine, and the timescale of the magneto-rotational instability. It can be seen, that duration of the pulses in the jet, which reveals the size and speed of blobs containing high thermal and Poynting energy, corresponds to the timescale of the fastest growing mode of MRI (cf. Fig. \ref{fig_mu}). Furthermore, we use the average duration of the pulses, measured at their half-width, as a proxy for the MTS. There is an anti-correlation found between this MTS proxy, and the black hole spin parameter of the central engine. The latter is directly responsible for the jest launching via the Blandford-Znajek process, so that the jet Lorentz factor will increase with the black hole spin, while the MTS timescale decreases with it.
Thus, the observed anti-correlation between MTS and $\Gamma$ is reproduced by our model (cf. \cite{Wu2016}). 

In addition, the MTS-$T_{90}$ correlation should be naturally reproduced. However, this is mainly due to the fact that the 
calculations are done in dimensionless unit system. Therefore, the simulations we run in dimensionless time units, $t_{g}=GM_{\rm BH}/c^{3}$, should be converted to physical timescales, assuming a fixed black hole mass.
The time unit for black hole of 10 Solar mass will be equal to $4.96 \times 10^{-5}$ s. The minimum variability timescales are for this conversion unit in between 1 and 2 ms, while the timescale of operation of the engine which we cover in our simulation, is on the order of 0.15 s (it has to be noted, that we are not running the models for a longer time, because of magnetic field decay and inefficient MRI turbulence at late times, which limits the effective accretion period, while the massive torus is still present and does not replenish, so the engine operation could last $\sim 100$ times longer). Therefore, adopting a range of black hole masses diving the central engine of a GRB, from $\sim3$ up to $\sim 30 M_{\odot}$, we will automatically be able to cover the range of $T_{90}$ duration times and MTS timescales in a correlated way. The scatter in this relation will be imposed by the additional factors, such as the mass of the disk available for accretion, and its magnetisation, hence the accretion rate. Furthermore, we can speculate that the relation between $\Gamma$ and MTS which spans $\sim 10$ orders of magnitude in the observations presented by \citet{Wu2016}, can also reach the blazar sample. The black hole mass in our simulations scales the MTS via the gravitational time scale, up to $\log(T)\sim 5$ for a black hole mass of $10^{8} M_{\odot}$. The smaller values of Lorentz factor should be related mainly with smaller black hole spin parameter.

We notice that our MTS (measured on average within the jet) is affected by the magnetic field strength, but assuming a given black holes spin we can have either the shortest timescales for most magnetized tori (i.e. $a=0.99$, cf. Fig. \ref{correlations}), or the opposite ($a=0.7-0.8$). Therefore we conclude that it is the total efficiency of the Blandford-Znajek process, rather than single parameter of the engine, which drives the jet variability timescales.
Its observed value is further regulated by the factors describing efficiency of conversion of the jet bulk kinetic energy into radiation \citep{Granot2015}
which is beyond the scope of our present simulations.

\acknowledgements
We thank Kostas Sapountzis and Mariusz Tarnopolski for helpful discussions. We also thank the anonymous referee for many valuable suggestions that greatly improved our manuscript.
We acknowledge support by the grants no. 2016/23/B/ST9/03114 and 2019/35/B/ST9/04000 from the Polish National
Science Center, and acknowledges computational resources of the Warsaw ICM through grant g85-947, and the PL-Grid through the grant grb4.

\bibliography{grbcorr.bib}

\end{document}